\documentclass[a4paper,10pt]{article}
\pdfoutput=1
\usepackage{jheppub}
\usepackage[T1]{fontenc} 
\usepackage{float,tikz-cd,braket,datetime2,graphicx,color,amsfonts,amsthm,bm,bbm,mathrsfs,amsmath,float,pgf,tikz,mathrsfs,xcolor}
\usetikzlibrary{cd}
\definecolor{rindou1}{rgb}{0.4431,0.2862,0.7960}
\definecolor{rindou2}{rgb}{0.0078,0.1215,0.4392}
\definecolor{lapis}{rgb}{0.0.0470,0.2941,0.5568}
\definecolor{burgundy}{rgb}{0.5, 0.0, 0.13}
\usepackage[T1]{fontenc}
\usepackage[english]{babel}
\usepackage[utf8]{inputenc}
\usetikzlibrary{arrows}

\newcommand{\z}{\overline{z}}

\newcommand{\ab}[1]{\bigg\langle{#1\bigg\rangle}}

\title{The Holographic Soft $S$-Matrix in QED and Gravity}

\author{Nikhil Kalyanapuram}

\affiliation{Department of Physics and Institute for Gravitation and the Cosmos, The Pennsylvania State University, University Park PA 16802, USA}

\emailAdd{nkalyanapuram@psu.edu}

\abstract{At leading order, the $S$-matrices in QED and gravity are known to factorise, providing unambiguous determinations of the parts divergent due to infrared contributions. The soft $S$-matrices defined in this fashion are shown to be defined entirely in terms of $2$ dimensional models on the celestial sphere, involving two real scalar fields, allowing us to express the soft $S$-matrices for real as well as virtual divergences as two dimensional correlation functions. We discuss what this means for finding holographic representations of scattering amplitudes in QED and gravity and comment on simple double copy structures that arise during the analysis.}

\begin{document} 
\maketitle
\flushbottom

\section{Introduction}\label{sec:intro}
Historically, the concerns with divergences in the $S$-matrix for quantum field theories of interest revolved around complications deep in the ultraviolet. However, it has been known for a long time that $S$-matrices for theories having massless particles exhibit singularities not just in the UV, but also in the infrared domain, where interacting particles approach arbitrarily low frequencies. Practically, such divergences were known to cancel when final state particles were properly summed over \cite{PhysRev.52.54,Jauch:1976ava,Yennie:1961ad,Kinoshita:1962ur,Lee:1964is}. Accordingly, it was believed for a long time that such divergences were essentially spurious\footnote{Actually, this exact word was used by Schwinger in his seminal paper on quantum electrodynamics in 1949 \cite{PhysRev.76.790}.}. However, more sophisticated treatments due to Chung \cite{Chung:1965zza} and Kibble \cite{Kibble:1968sfb,Kibble:1969ip,Kibble:1969ep,Kibble:1969kd} followed, in which IR divergences were eliminated by dressing asymptotic states. 

In \cite{Weinberg:1965nx}, a seminal paper due to Weinberg, it was shown that not unlike QED, gravity amplitudes display very simple analytic forms deep in the infrared - the cancellation of the concomitant divergences proceeded by including final state radiation\footnote{This result was decisively established by DeWitt in \cite{DeWitt:1967uc}.}. 

Fundamentally, divergences deep in the IR are possible because of the fact that both in QED and gravity, the fundamental particle mediating interactions, namely the photon and graviton respectively, is massless. Consequently, they give rise to interactions which are long ranged, a fact that manifests itself as IR divergences in the $S$-matrix. From the quantum field theoretic perspective however, most attention remained fixed on understanding IR divergences as analytic phenomena, rather than as something deeper about the structure of the theory.

Nearly simultaneously, it was shown by Penrose \cite{Penrose:1965am} and by Newman and Penrose \cite{Newman:1968uj} that the asymptotic structure of theories of QED and gravity is really quite rich, precisely as a consequence of the fact that they
evolve in a nontrivial fashion on null infinity. The asymptotic behaviour of gravity was especially revealing as careful analyses \cite{PhysRev.128.2851,Bondi:1962px} showed that in the presence of gravity, the metrics of spacetimes which are asymptotically flat are no longer invariant only under the Poincar\'e group - the group is enhanced to an infinite dimensional group now known as the Bondi-Metzner-Sachs (BMS) group . 

The natural question of whether or not these seemingly disparate ways of thinking about the long distance effects of QED and gravity are related has received attention in recent years (see \cite{Strominger:2017zoo} and references therein). As it turns out, the soft graviton theorem due to Weinberg, which tells us what happens to the $S$-matrix when soft gravitons are radiated in a scattering process, turns out to be equivalent to the statement that the $S$-matrix remain invariant under the action of supertranslations belonging to the BMS group. 

What does the relationship between asymptotic symmetries like BMS symmetry and soft theorems tell us? For one thing, it tells us that the analytic structure of the $S$-matrix in the infrared - something that appears to be dependent on the nature of interactions in the bulk - depends on data that is present entirely on the boundary of four dimensional spacetime, namely null infinity in the form on BMS charges. In particular, it suggests the possibility of realising the entire $S$-matrix, at least in fundamentally massless theories like QED and gravity, in terms of purely boundary data - a holographic representation, in other words.

Needless to say, such a framework is probably not easy to find - not least due to the fact that gravity in particular is extremely hard to work with. Accordingly, a more modest approach would be to wonder if partial progress can be made. The natural first step of course would be to deal directly with the soft part of the $S$-matrix and ask if a truly holographic dynamical framework which captures soft theorems can be found, which makes absolutely no reference to anything in the bulk. We have attempted in this article to provide an affirmative response to this question. To do so, we leverage two ideas which are already in the literature, which we will now discuss.

The first is a phenomenon known as factorisation \cite{Collins:1988ig,Collins:1989bt,Collins:1989gx,Feige:2013zla,Feige:2014wja,Akhoury:2011kq} which is known to apply to a large class of massless quantum field theories, including QED and gravity. While we provide a brief review of this in the next section, we simply state at this point that according to factorisation theorems, the soft divergences in QED and gravity can be collected into universal expressions which form a part of any scattering amplitude in these theories, and can be \emph{factored out}, so to speak. In carrying out this factorisation, we obtain an unambiguous definition of the soft $S$-matrix, and can then ask whether these objects admit representations in terms of dynamical fields defined on the boundary.

What we will see in this article is that this is indeed possible in QED and gravity, in which the soft $S$-matrices have especially simple forms. Specifically, we ask whether it is possible to generalise the central results of \cite{Kalyanapuram:2020epb}. In that work, we saw that the soft $S$-matrix due to \emph{virtual} soft interactions in QED can be obtained as a correlation function of vertex operators in a Coulomb gas, defined naturally on the celestial sphere $\mathbb{CP}^{1}$. For the case of gravity, the holographic theory generalised very naturally, and was given in terms of a free scalar with a biharmonic kinetic term, with the soft $S$-matrix due to virtual transitions once again obtained as a correlation function of vertex operators. One drawback of that approach however was that it was difficult to include the effects of multiple real soft emissions. In this work, we will show that it is relatively straightforward to modify the formalism of \cite{Kalyanapuram:2020epb} to include multiple soft emissions as well.

Let us comment on the notation we will use before we move on. We work throughout with theories having only massless particles. Momenta $p_{i}$ are decomposed as

\begin{equation}
    p_{i} = \omega_{i}(1+z_{i}\z_{i},z_{i}+\z_{i},-i(z_{i}-\z_{i}),1-z_{i}\z_{i})
\end{equation}
where $\omega_{i}$ is the particle energy and the $z_{i}$ and $\z_{i}$ take values on $\mathbb{CP}^{1}$. A more convenient system of coordinates which we will employ throughout the paper is given by

\begin{equation}
    \begin{aligned}
    p^{+}_{i} &= p^{0}_{i}+p^{3}_{i},\\
    p^{-}_{i} &= p^{0}_{i}-p^{3}_{i},\\
    p^{z}_{i} &= p^{1}_{i}+ip^{2}_{i},\\
    p^{\z}_{i} &= p^{1}_{i}-ip^{2}_{i},\\
    \end{aligned}
\end{equation}
so that the momentum is expanded according to

\begin{equation}
    p_{i} = \omega_{i}(1,z_{i}\z_{i},z_{i},\z_{i}).
\end{equation}

This paper has been organised to be relatively self-contained; in section \ref{sec:soft} we review the factorisation theorems in gauge theory and gravity, which make possible precise definitions of soft $S$-matrices in the two theories. This is followed by a proposal for the holographic description of the soft $S$-matrices in QED and gravity in section \ref{sec:hol}. How this proposal ties in with the broader program of finding holographic representations of the entire theories is considered and discussed in section \ref{sec:qedgravity} after which the paper concludes with a summary and future directions in section \ref{sec:discussion}.

\section{Soft $S$-Matrices from Factorisation}\label{sec:soft}
Before trying to develop two dimension models that encode the soft part of the $S$-matrices for QED and gravity, it is helpful to have an unambiguous presentation of what it really means to compute the soft $S$-matrix. We work at leading order throughout henceforth.

In the case of QED, in addition to infinities in the ultraviolet there are two classes of divergences that one encounters. The first, which is what we will consider in this paper is in the infrared sector - where one or more photons is taken to be deep in the infrared, namely that it's energy approaches zero. The second class of divergences is due to to collinear interactions, in which a number of photons approach a given external matter leg. 

To take a concrete example, one may think of an external matter state like an electron. In high energy scattering the electron is essentially massless. Accordingly the direction of its momentum is simply a point on the celestial sphere. When one or more external photon states, which may or may not be massless approach(es) has momentum almost parallel to that of this electron, the $S$-matrix becomes singular. The statement of factorisation is simply that the $S$-matrix factorises into a product of collinear contributions, corresponding to points separated on the celestial sphere each of which consists of a number of collinear states and two soft pieces, which arise out of virtual and real soft emissions. Importantly, the soft and collinear interactions decouple and provide independent sources of divergences. Indeed, we have for a QED $S$-matrix element \cite{Collins:1988ig,Collins:1989bt,Collins:1989gx,Feige:2013zla,Feige:2014wja}

\begin{equation}
    \mathcal{M}^{(1)}_{n} = \prod_{n_{i}}\mathcal{C}_{n_{i}}\times\mathcal{A}^{soft}_{n,s=1}|_{vir}\times \mathcal{A}^{soft}_{n,s=1}|_{real}\times \mathcal{H}_{n,s=1}.
\end{equation}
Here, each $\mathcal{C}_{n_{i}}$ encodes collinear interactions of a given sector, where the $n_{i}$ is used to indicate the number of external states in that sector. The function $\mathcal{A}^{soft}_{n,s=1}|_{vir}$ is the soft $S$-matrix due to virtual soft photon transitions, and is given by

\begin{equation}
    \ln\left(\mathcal{A}^{soft}_{n,s=1}|_{vir}\right) = -\frac{1}{8\pi^{2}\epsilon}\sum_{i\neq j}e_{i}e_{j}\ln|z_{i}-z_{j}|^{2}
\end{equation}
where the $e_{i}$ are charges of the external states and the $z_{i}$ are punctures on the celestial sphere coming from the momenta of the external states. The function $\mathcal{A}^{soft}_{n,s=1}|_{real}$ is the contribution to the $S$-matrix arising out of the emission of real soft photons. It depends on the number of soft photons emitted. Given the soft factor for the emission of a positive helicity photon

\begin{equation}
    \mathcal{S}^{(1)}_{0}(z) = \frac{i}{\pi}\sum_{i}\frac{e_{i}}{z-z_{i}}.
\end{equation}
the object $\mathcal{A}^{soft}_{n,s=1}|_{real}$ for one soft positive helicity photon emission along the direction given by $z$ with energy $\omega$ is given by

\begin{equation}
    \mathcal{A}^{soft}_{n,s=1}|_{real} = \frac{1}{\omega}S^{(1)}_{0}(z).
\end{equation}
When $N$ such soft photons are radiated, $\mathcal{A}^{soft}_{n,s=1}|_{real}$ is composed of $N$ soft factors. When negative helicity photons are involved, the complex conjugated version of $\mathcal{S}^{(1)}_{0}(z)$ encodes such an emission.

Finally, the function $\mathcal{H}_{n,s=1}$ is the so-called hard $S$-matrix, which has no infrared or collinear divergences. It is believed that it can be defined in an intrinsic fashion - see for example \cite{Hannesdottir:2019rqq,Hannesdottir:2019opa}. 

The case of gravity is surprisingly much simpler at leading order. A key fact about factorisation of gravity amplitudes at leading order is that there are \emph{no collinear divergences} in gravity. This statement was known to be true already by Weinberg \cite{Weinberg:1965nx} and DeWitt \cite{DeWitt:1967uc}, has now been established in more modern settings as well \cite{Akhoury:2011kq}. The theorem of factorisation then boils down to a very simple statement about matrix elements $\mathcal{M}^{(2)}_{n}$ in a gravitational theory:

\begin{equation}
    \mathcal{M}^{(2)}_{n} = \mathcal{A}^{soft}_{n,s=2}|_{vir}\times \mathcal{A}^{soft}_{n,s=2}|_{real}\times \mathcal{H}_{n,s=2}. 
\end{equation}
This time, the object $\mathcal{A}^{soft}_{n,s=2}|_{vir}$ is defined according to

\begin{equation}
    \ln\left(\mathcal{A}^{soft}_{n,s=2}|_{vir}\right) = -\frac{\kappa^{2}}{8\pi^{2}\epsilon}\sum_{i<j}\omega_{i}\omega_{j}|z_{i}-z_{j}|^{2}\ln|z_{i}-z_{j}|^{2}
\end{equation}
where $\kappa^{2} = 8\pi G$. The $\omega_{i}$ have been used to denote the energies of the external states.

The soft factor is gravity takes a form that is very similar to that in QED. Indeed, for a positive helicity graviton we have

\begin{equation}
     \mathcal{S}^{(2)}_{0}(z) = \frac{i}{\pi}\sum_{i}\omega_{i}\kappa\frac{\z-\z_{i}}{z-z_{i}}.
\end{equation}
Accordingly, the soft $S$-matrix in gravity due to one real emission of a soft positive helicity graviton is given by 

\begin{equation}
    \mathcal{A}^{soft}_{n,s=2}|_{real} = \frac{1}{\omega}\mathcal{S}^{(2)}_{0}(z)
\end{equation}
with an analogous expression for a negative helicity emission. Multiple soft emissions supply a corresponding number of such factors. The hard $S$-matrix in gravity has of course been denoted by $\mathcal{H}_{n,s=2}$, which is found by simply computing the $S$-matrix element in the absence of any soft corrections. 

In this article, we will be mainly interested in the soft contributions we have just discussed. As the reader will observe, since we are dealing with massless external states, the soft $S$-matrices are defined entirely in terms of coordinates on the celestial sphere, which label directions of the external states. We can then ask if there is some more fundamental way to derive these expressions directly from theories defined on the celestial sphere. Put differently, are there two dimensional quantum field theories which naturally compute both corrections to the $S$-matrices in QED and gravity due to real and virtual soft interactions? As we will see, there exist two dimensional models which do precisely this.

\section{The Holographic Models for the Soft $S$-Matrices}\label{sec:hol}
We start by considering the following two dimensional theory of scalar fields on $\mathbb{CP}^{1}$

\begin{equation}
S^{(1)}=\int d^{2} z\left[g_{a b} \partial_{i} \varphi^{a} \partial^{i} \varphi^{b}\right].
\end{equation}
Since the theory is defined in two dimensions, the partial derivatives $\partial_{i}$ run over a two dimensional coordinate system on the celestial sphere given in terms of the pair $(z,\overline{z})$. The indices $a$ and $b$ take the values $\lbrace{1,2\rbrace}$. Without any loss of generality of course, the metric $g_{ab}$ is taken to be symmetric.

For the right choice of the metric, a convenient operator product expansion can be engineered between the scalar fields. We make the particular choice

\begin{equation}
g_{a b}=\left[\begin{array}{rr}
-a & 1 \\
1 & 0
\end{array}\right].
\end{equation}
Obviously, this is different from a normal theory of two scalar fields, in which the fields may be rotated by an $O(2)$ rotation while keeping the theory invariant. Our choice of $g_{ab}$ explicitly breaks this symmetry.

Given this information, the operator product expansion between two scalar fields can be inferred by inspection of the correlation function

\begin{equation}
\left\langle\varphi^{a}(z) \varphi^{b}\left(z^{\prime}\right)\right\rangle=\left(g^{-1}\right)^{a b} \frac{1}{\pi} \ln \left|z-z^{\prime}\right|^{2}
\end{equation}
More explicit is the expansion in terms of the component fields:

\begin{equation}
\begin{aligned}
\left\langle\varphi^{1}(z) \varphi^{1}\left(z^{\prime}\right)\right\rangle&=0\\
\left\langle\varphi^{1}(z) \varphi^{2}\left(z^{\prime}\right)\right\rangle&=\frac{1}{\pi} \ln \left|z-z^{\prime}\right|^{2} \\
\left\langle\varphi^{2}(z) \varphi^{2}\left(z^{\prime}\right)\right\rangle&=\frac{a}{\pi} \ln \left|z-z^{\prime}\right|^{2}. 
\end{aligned}
\end{equation}
Before moving on to specific considerations of correlation functions of such fields, it is useful to determine and characterise the nature of the global symmetries enjoyed by this model. Like the ordinary Coulomb gas, this two dimensional model exhibits a shift symmetry \cite{DiFrancesco:639405}, according to the replacements

\begin{equation}
    \varphi^{a} \rightarrow \varphi^{a} + c^{a}
\end{equation}
where the vector $c^{a}$ is just composed of two constants, which are independent. As it turns out, for the purposes of this work we will not have occasion to exploit this full symmetry, but will rather only be interested in the shift symmetry corresponding to the second field:

\begin{equation}
    \varphi^{2} \rightarrow \varphi^{2} +c
\end{equation}
where we have chosen to drop the index on the constant. Since this is a global symmetry, we are given our choice of holomorphic and antiholomorphic currents that give rise to the conserved charge corresponding to the global transformation given above. Indeed, we have the following two currents

\begin{equation}\label{eq:3.7}
\begin{aligned}
j^{(1)}(\z)&=\overline{\partial} \varphi^{1}(z, \overline{z}) \\
\overline{j}^{(1)}(z)&=\partial \varphi^{1}(z, \overline{z}).
\end{aligned}
\end{equation}

Reproducing the soft $S$-matrix due to virtual corrections is now a matter of noting that the vertex operators $\mathcal{V}^{(1)}(z_{i})$ defined according to

\begin{equation}
    \mathcal{V}^{(1)}(z_{i},\z_{i}) = \exp\left(ie_{i}\varphi^{2}(z_{i},\z_{i})\right)
\end{equation}
can be used to show that

\begin{equation}\label{eq:3.9}
    \ab{\mathcal{V}^{(1)}(z_{1},\z_{1})\dots \mathcal{V}^{(1)}(z_{n},\z_{n})} = \exp\left(-\frac{a}{\pi}\sum_{i<j}e_{i}e_{j}\ln|z_{ij}|^{2}\right)
\end{equation}
where $z_{ij} = z_{i}-z_{j}$. In accordance with this, it is clear that a convenient choice of $a$ will give the right soft factor. Indeed, we can define an $a$ to suit us both in dimensional regularization (in $4+2\epsilon$ dimensions) as well as in infrared regularization. For the former, the right choice of $a$ is given by

\begin{equation}
    a_{DR} = \frac{1}{8\pi\epsilon}
\end{equation}
while for the latter we need to choose

\begin{equation}
    a_{IR} = -\frac{1}{8\pi}\ln\left(\frac{\Lambda_{UV}}{\lambda_{IR}}\right)
\end{equation}
where $\Lambda_{UV}$ is some RG scale with $\lambda_{IR}$ is a choice of IR cut off. Although these are divergent, since IR finite observables are defined (think for example of the ratio function in $\mathcal{N}=4$ super Yang-Mills theory) as ratios of $S$-matrices, the divergent nature of the constant $a$ poses no problem to defining IR safe quantities.

In \cite{Kalyanapuram:2020epb} a similar model was proposed to handle only the virtual divergences, and currents entirely analogous to those in (\ref{eq:3.7}) were used to show that the conservation law corresponding to them was simply the statement of charge conservation, arising out of the associated real soft theorem. However, that model suffered from an ambiguity in that two such insertions did not produce the double soft theorem. The advantage of the model proposed here is that we can circumvent this problem.

Consider again the insertion of a single holomorphic current $j^{(1)}(z)$ into the correlation function in (\ref{eq:3.9}). Employing the operator product expansions derived earlier, we have 

\begin{equation}\label{eq:3.12}
    \ab{j^{(1)}(z)\mathcal{V}^{(1)}(z_{1})\dots \mathcal{V}^{(1)}(z_{n})} = \left(\frac{i}{\pi}\sum_{i}\frac{e_{i}}{z-z_{i}}\right)\exp\left(-\frac{a}{\pi}\sum_{i<j}e_{i}e_{j}\ln|z_{ij}|^{2}\right)
\end{equation}
which is obviously the leading order soft theorem corresponding to the emission of a single soft photon. Noting of course that

\begin{equation}
    \mathcal{S}^{(1)}_{0}(z) = \frac{i}{\pi}\sum_{i}\frac{e_{i}}{z-z_{i}}.
\end{equation}
the double soft emission theorem is then obtained by simply inserting two soft currents to give

\begin{equation}\label{eq:3.14}
    \ab{j^{(1)}(z')j^{(1)}(z)\mathcal{V}^{(1)}(z_{1})\dots \mathcal{V}^{(1)}(z_{n})} =  \mathcal{S}^{(1)}_{0}(z') \mathcal{S}^{(1)}_{0}(z)\exp\left(-\frac{a}{\pi}\sum_{i<j}e_{i}e_{j}\ln|z_{ij}|^{2}\right).
\end{equation}
Naturally, this is easy to generalise to any number of soft insertions. Inserting $N$ soft currents leads to the soft theorem corresponding to the emission of $N$ soft photons. 

We can review how charge conservation fits into this formalism. Indeed, notice that each of the scalar fields in the correlator (\ref{eq:3.9}) can be shifted by a constant $c$ according to the global symmetry of the model. Insisting on the invariance of the correlator under this transformation requires that the phase

\begin{equation}
    \delta_{1} = i\sum_{i}e_{i}
\end{equation}
vanishes. This is just charge conservation. Alternatively, this can be derived by applying the global residue theorem to the holomorphic soft current, which gives us precisely the preceding statement. Parenthetically, we observe that the exact same calculation goes through for the antiholomorphic current. The holomorphic and antiholomorphic currents simply corresponding to emissions of soft photons of positive and negative helicity respectively.

We point out at this stage that we have - using this model - resolved the soft part of the $S$-matrix in QED completely from the hard part, so long as the factorisation theorem at leading order is assumed. Importantly, this model provides a complete realisation of the soft part of the QED $S$-matrix in terms of a two dimensional theory, which is precisely what one wants in a holographic description. While this is interesting, holography is most relevant in the context of gravity. We turn now to generalising the model we have just described to the case of the soft $S$-matrix in gravity.

Informed by the calculations performed in \cite{Kalyanapuram:2020epb}, let us consider the following model on the celestial sphere

\begin{equation}\label{eq:3.16}
S^{(2)}=\int d^{2} z\left[g_{a b} \partial_{i}\partial_{j} \chi^{a} \partial^{i}\partial^{j} \chi^{b}\right].
\end{equation}
As pointed out in \cite{Kalyanapuram:2020epb}, this is a theory of two dimensional disclinations. Although it looks like it's a theory of two different flavours of disclinations, due to the basic arguments of the preceding discussion, disclinations of only one variety are ultimately involved. Noting once again that the Green's function of the biharmonic operator $\nabla^{2}$ is just $\frac{1}{\pi}|z|^{2}\ln|z|^{2}$, we have the following OPEs

\begin{equation}
\begin{aligned}
\left\langle\chi^{1}(z) \chi^{1}\left(z^{\prime}\right)\right\rangle&=0\\
\left\langle\chi^{1}(z) \chi^{2}\left(z^{\prime}\right)\right\rangle&=\frac{1}{\pi} \left|z-z^{\prime}\right|^{2}\ln \left|z-z^{\prime}\right|^{2} \\
\left\langle\chi^{2}(z) \chi^{2}\left(z^{\prime}\right)\right\rangle&=\frac{a}{\pi} \left|z-z^{\prime}\right|^{2}\ln \left|z-z^{\prime}\right|^{2}. 
\end{aligned}
\end{equation}

The global symmetries of this theory are inferred by noting that the action in (\ref{eq:3.16}) is either invariant or changes by a total derivative under the transformation

\begin{equation}
    \chi^{a}(z,\z) \rightarrow \chi^{a}(z,\z) + c^{a}_{1} + c^{a}_{2}z + c^{a}_{3}\z + c^{a}_{4}z\z.
\end{equation}
Since the constants $c^{a}_{i}$ are arbitrary, the global symmetry is an $8$ parameter group.

In complete analogy to the results of the earlier discussion, we note the following vertex operator

\begin{equation}
    \mathcal{V}^{(2)}(z_{i},\z_{i}) = \exp\left(i\omega_{i}\kappa_{i}\chi^{2}(z_{i},\z_{i})\right)
\end{equation}
where $\omega_{i}$ is to be identified with the energy of an external state while $\kappa_{i}$ is some constant, that \emph{a priori} is label dependent, although (as we expect) will be shown to be universal in accordance with the equivalence principle. In a nod to this fact, we will need the following currents

\begin{equation}\label{eq:3.20}
\begin{aligned}
j^{(2)}(z,\z)&= \partial^{2}\chi^{1}(z, \overline{z}) \\
\overline{j}^{(2)}(z,\z)&=\overline{\partial}^{2} \chi^{1}(z, \overline{z}).
\end{aligned}
\end{equation}

Moving now to showing that we can find the soft $S$-matrix for gravity in this model, we simply compute the following correlator, which immediately gives the desired result

\begin{equation}\label{eq:3.21}
    \ab{\mathcal{V}^{(2)}(z_{1},\z_{1})\dots \mathcal{V}^{(2)}(z_{n},\z_{n})} = \exp\left(-\frac{a}{\pi}\sum_{i<j}\omega_{i}\omega_{j}\kappa_{i}\kappa_{j}|z_{ij}^{2}|\ln|z_{ij}|^{2}\right)
\end{equation}
in precise agreement with the virtual soft factor in gravity \cite{Weinberg:1965nx}. Before moving on to the soft theorem, we see that the global symmetry of this model introduces the following phase into the correlator
\begin{equation}
    \delta_{2} = i\sum_{i}\omega_{i}\kappa_{i} + i\sum_{i}\omega_{i}\kappa_{i}z_{i} + i\sum_{i}\omega_{i}\kappa_{i}\z_{i} + i\sum_{i}\omega_{i}\kappa_{i}z_{i}\z_{i}. 
\end{equation}
That this must vanish is simply the statement that

\begin{equation}
    \sum_{i}\kappa_{i}p^{\mu}_{i} = 0
\end{equation}
which in combination with momentum conservation tells us that the constants $\kappa_{i}$ must all equal some constant $\kappa$. This is just the celestial analogue of Weinberg's argument in \cite{Weinberg:1964ew}.

So far what we have done is not altogether different from what we showed in \cite{Kalyanapuram:2020epb}. Once again, the difference is in the real sector. First, we observe the following

\begin{equation}
    \langle{j^{(2)}(z,\z)\mathcal{V}^{(2)}(z_{i},\z_{i})\rangle} = \frac{i\omega_{i}\kappa}{\pi}\frac{\z-\z_{i}}{z-z_{i}}\mathcal{V}^{(2)}(z_{i},\z_{i}).
\end{equation}
This tells us that a single insertion of the soft current reproduces the soft theorem due to a single soft graviton emission of positive helicity:

\begin{equation}\label{eq:3.25}
    \ab{j^{(2)}(z,\z)\mathcal{V}^{(2)}(z_{1},\z_{1})\dots \mathcal{V}^{(2)}(z_{n},\z_{n})} = \mathcal{S}^{(2)}_{0}(z)\exp\left(-\frac{a}{\pi}\sum_{i<j}\omega_{i}\omega_{j}\kappa_{i}\kappa_{j}|z_{ij}^{2}|\ln|z_{ij}|^{2}\right).
\end{equation}
Obviously, the same calculation goes through for the current $\overline{j}^{(2)}(z,\z)$, which corresponds to the emission of a negative helicity graviton. There are once again no nontrivial OPEs between the currents themselves, and inserting $N$ such currents gives the soft theorem for $N$ soft graviton emissions.

It is clear from this discussion that what we have is a genuine holographic theory for the soft $S$-matrix in gravity. Importantly, the results in this section have been derived by making no reference to fields in the bulk and are defined intrinsically on null infinity. 

\section{Towards a Holographic Model of QED and Gravity in Flat Space}\label{sec:qedgravity}
One reason for the interest in expressing scattering amplitudes intrinsically on the celestial sphere is hoping that doing it might supply a path towards further understanding instances of flat space holography. Given the results of the previous section, it is natural to wonder how the two dimensional representations of the soft $S$-matrices in QED and gravity affect our ability to construct holographic versions of the full theories. As it turns out, the cases of QED and gravity differ somewhat, so we discuss them in turn.

If we look at the QED case, the first thing to observe is that the soft factors in QED are energy independent, in that they do not depend on the energies of the external states involved in the scattering process. Why is this helpful in developing a holographic dual that can compute scattering amplitudes in QED? To understand this, we turn our attention to one proposal for constructing amplitudes that at least on the surface appear to have holographic origins, given amplitudes computed in a standard fashion from say QED.

Suppose we have a scattering amplitude, which encodes maybe a number of collinear sectors, in QED which we denote by $\mathcal{M}^{(1)}_{n}$. In computing such amplitudes, momentum conservation is ensured by constructing asymptotic states that are in a momentum eigenbasis. Due to this fact, at least when we are working with massless amplitudes, the scattering amplitude will depend on the energies $\omega_{i}$ of the external states along with the coordinates $(z_{i},\z_{i})$, which represent punctures on the celestial sphere. One way to construct, by hand, a holographic representation of such amplitudes is to find a basis which exhibits the right behaviour under \emph{conformal transformations}, which is what we expect for a holographic theory. It turns out that such a basis is furnished by the so-called Mellin basis, which converts momentum eigenstates into boost eigenstates via the transform\footnote{The transform we have decided to discuss is known as the modified Mellin transform, due to Banerjee \cite{Banerjee:2018gce,Banerjee:2018fgd,Banerjee:2019prz}. It is a refinement of the ordinary Mellin transform due to Pasterski et al. \cite{Pasterski:2016qvg,Pasterski:2017ylz}, which does not converge for scattering amplitudes that are not UV complete.}

\begin{equation}
    \widetilde{\mathcal{M}}^{(1)}_{n} = \int \prod_{i=1}^{n} \omega_{i}^{i\Lambda_{i}}e^{iu_{i}\omega_{i}} d\omega_{i} \times \mathcal{M}^{(1)}_{n}
\end{equation}
where the integration is carried out over the energies of the external states. This transform naturally yields a scattering amplitude that is intrinsically defined on $\mathbb{R}\times\mathbb{CP}^{1}$, namely on future null infinity, where the retarded coordinate is labelled by $u$. 

This is where the fact that the soft part of the QED $S$-matrix being independent of the energies of the external states becomes relevant. Indeed, by virtue of this fact the soft factorisation \emph{continues to hold} even at the level of the celestial amplitudes obtained from the Mellin transform. This tells us that insofar as there exists a holographic description of the collinear and hard sectors of QED (which is by no means certain), the full holographic theory decomposes very nicely - into a soft part which is described by an $O(2)$-broken Coulomb gas model, and a putative holographic model that encodes the collinear and hard parts.

Things become more interesting (and less simple) when we want to discuss the prospect of constructing a holographic model of gravity in flat space. First, we point out that due to the absence of collinear interactions at leading order in factorisation, an amplitude in gravity is broken up into a hard part and soft corrections - and we know that the latter is fully described by an $O(2)$-broken model of crystal disclinations. This being said, let us see why the inclusion of soft modes in gravity is a considerable complicating factor. Consider an $n$-point function for graviton scattering, which we denote by $\mathcal{M}^{(2)}_{n}$. In the absence of soft interactions, the corresponding amplitude on null infinity is obtained by the modified Mellin transform

\begin{equation}
    \widetilde{\mathcal{M}}^{(2)}_{n} = \int \prod_{i=1}^{n} \omega_{i}^{i\Lambda_{i}}e^{iu_{i}\omega_{i}} d\omega_{i} \times \mathcal{M}^{(2)}_{n}.
\end{equation}
What happens now if we want to include soft interactions? In the QED case, since the vertex operators had no $\omega_{i}$ dependence, the inclusion of soft modes simply devolved upon multiplying the Mellin amplitude by the soft $S$-matrix. In the case of gravity however, precisely due to the fact that the vertex operators have $\omega_{i}$ dependence, the scattering amplitude in momentum space must be dressed by these factors \emph{before} carrying out the Mellin transform. The result is as follows:

\begin{equation}
    \widetilde{\mathcal{M}}^{(2)}_{n} = \int \prod_{i=1}^{n} \omega_{i}^{i\Lambda_{i}}e^{i\omega_{i}(u_{i}+\chi(z_{i},\z_{i}))} d\omega_{i} \times \mathcal{M}^{(2)}_{n}.
\end{equation}
Two facts can be inferred from the preceding expression. Most obviously, factorisation is clearly lost when we move to the Mellin basis in the case of gravity. This in itself is not a problem - it just tells us that in coming up with a holographic theory of gravity, even at tree level and leading order factorisation, it will likely be important to see how the soft and hard parts interact, since it is clear that they do not decouple in the celestial basis.

Amusingly, the inclusion of the soft mode has another effect - it shifts the location of the retarded coordinate $u_{i}$ of the $i$th external state according to

\begin{equation}
    u_{i} \rightarrow u_{i} + \chi(z_{i},\z_{i}).
\end{equation}
Classically, what this tells us is that the inclusion of soft modes has the effect of supertranslating the retarded coordinate by a biharmonic function on the celestial sphere. Of course, one could have seen this even at the level of the ordinary Mellin transform, which is obtained by simply setting all the $u_{i}$ to vanish, but the nature of the shift is most clearly expressed in the modified Mellin basis.

In closing, the holographic dual description of the soft $S$-matrix in QED is valuable in that factorisation at leading order is preserved by the Mellin transform, meaning that finding a holographic form of the full theory boils down to constructing such a dual for the collinear and hard sectors separately. As we have come to expect however, gravity is not as easy to work with as the two dimensional model of soft divergences in gravity will probably mix with the hard part upon going to a celestial basis\footnote{We mention parenthetically the papers \cite{Adamo:2014yya,Adamo:2015fwa,Adamo:2019ipt}, in which progress was made in developing truly two dimensional models (in that they are worldsheet theories) that attempt to describe the tree level dynamics of gauge theory and gravity. One possible direction of further study to realise holographic characterisations of these theories may be to place the models for soft dynamics suggested in this article consistently in the context such worldsheet models.}. 

\section{Discussion}\label{sec:discussion}
In this article we have constructed two dimensional models that capture the entire soft dynamics - due to real as well as virtual soft radiation - in QED and gravity. In particular, we have observed that we can construct models that are explicitly two dimensional, in that they are defined on the celestial sphere $\mathbb{CP}^{1}$, which then yield the soft parts of the $S$-matrices in QED and gravity directly as correlation functions of currents and vertex operators defined in terms of the fundamental fields of the theories. 

Specifically, the theories are those of two scalar fields - free scalar fields in the case of QED and scalar fields with a biharmonic kinetic term in the case of gravity. Unlike the conventional theory of two real scalars, the theories we deal with were \emph{not} invariant under global $O(2)$ transformations of the vector of scalars. Rather, the $O(2)$ symmetry was explicitly broken by a metric $g_{ab}$, which ensured that only one of the fields had a nontrivial OPE with itself. This made it possible to readily resolve the real and virtual parts of the soft $S$-matrix.

One aspect of the present analysis that requires some discussion is that of the double copy structure that has emerged. The double copy more generally is a phenomenon in which scattering amplitudes in a gravitational theory can be constructed by squaring corresponding amplitudes in some gauge theory. There is some evidence to suggest that the double copy is fundamentally a string-theoretic phenomenon \cite{Kawai:1985xq,Mizera:2017cqs,Mizera:2019gea,Mizera:2019blq,Kalyanapuram:2021xow,Kalyanapuram:2021vjt}, although more work needs to be done to establish whether or not this is indeed the case\footnote{The double copy seems also to be best expressed in momentum space. Progress was made in extending it to celestial amplitudes in \cite{Casali:2020vuy,Casali:2020uvr,Kalyanapuram:2020aya}}. At any rate, we saw in \cite{Kalyanapuram:2020epb} that soft $S$-matrices in QED and gravity enjoyed a double copy which was quite independent of any underlying string theoretic framework. Indeed, all that had to be done was to carry out the following replacements of the operators $\partial\overline{\partial}$ in the kinetic term

\begin{equation}
    \partial\overline{\partial} \rightarrow (\partial\overline{\partial})^{2}.
\end{equation}
In the present work, something entirely similar has happened. Indeed, the replacement given above has now been refined; one has to only carry out the following replacements

\begin{equation}
    \begin{aligned}
    \partial \rightarrow & \partial^{2}\\
    \overline{\partial} \rightarrow & \overline{\partial}^{2}\\
    \end{aligned}
\end{equation}
to move from the description of the QED soft $S$-matrix to the one in gravity. It is quite gratifying to see that not only is there such a simple squaring operation that takes us from one theory to the other, but also that it holds for the full soft $S$-matrix, not just for the virtual part as in \cite{Kalyanapuram:2020epb}.

One natural avenue of future research is to generalise the formalism presented here for QED to the more general problem of understanding the soft $S$-matrix in gauge theories. In \cite{Magnea:2021fvy,Gonzalez:2021dxw}, the authors have considered an action of the form

\begin{equation}
    S_{gauge} = \frac{1}{2\pi}\int d^{2}z[\partial_{i}\phi^{a}\partial^{i}\phi^{a}]
\end{equation}
where the scalars $\phi^{a}$ belong to a multiplet of the Lie algebra of the non-Abelian gauge theory. By defining vertex operators as

\begin{equation}
    \mathcal{V}^{(r)}_{gauge}(z,\z) = \exp\left(ikT^{a}_{r}\phi^{a}(z,\z)\right)
\end{equation}
where $k$ is some coupling constant and $T^{a}$ denote the generators of the representation labelled by $r$, the authors showed that the conformal correlation function of the vertex operators supplied the colour correlated part of the dipole contribution to the soft anomalous dimension. Once again, only virtual interactions were considered - extensions of the result to real corrections and nondipole corrections \cite{Almelid:2015jia,Almelid:2017qju,Grozin:2017css,Moch:2018wjh,Henn:2019rmi,Becher:2019avh,Dixon:2019lnw} remain open problems.

Finally, in \cite{Cheung:2016iub}, the Kac-Moody algebra that gave rise to soft theorems in gauge theory and the corresponding generalisation to gravity were related to three dimensional Chern-Simons models (the Chern-Simons model has shown up in other, unrelated studies of soft theorems as well in \cite{Balachandran:2013wsa,Balachandran:2014hra}). It would be intriguing to see whether or not the models we have studied here and in \cite{Kalyanapuram:2020epb} can be extracted as boundary theories of the models considered in \cite{Cheung:2016iub}.

\section*{Acknowledgements} 
I thank Jacob Bourjaily, Lorenzo Magnea, Sruthi Narayanan, Suvrat Raju and Vasudev Shyam for discussions. The research of the Bourjaily group at Penn State is supported in part by an ERC Starting Grant (No. 757978) and a grant from the Villum Fonden (No. 15369).



\bibliographystyle{utphys}
\bibliography{v1.bib}

\end{document}